\documentclass[aps,pra,showpacs,twocolumn,amsmath,amssymb,superscriptaddress,nofootinbib]{revtex4}
\usepackage{graphicx}% Include figure files
\usepackage{bm}% bold math
\usepackage{amssymb}
\usepackage{amsmath}
\usepackage{latexsym}
\usepackage{color}
\newcommand{\ind}[1]{_{#1}}    % generalized index instead of the specific
\begin{document}  
\title{Diamond Channel-Cut Crystals for High-Heat-Load, Beam-Multiplexing, Narrow-Band X-ray Monochromators}

\author{Yuri Shvyd'ko} \email{shvydko@anl.gov}
\affiliation{Advanced Photon Source, Argonne National Laboratory,  Argonne, Illinois 60439, USA}

\author{Sergey Terentyev}
\author{Vladimir Blank}
\affiliation{Technological Institute for Superhard and Novel Carbon  Materials,  142190~Troitsk, Russian Federation}

\author{Tomasz Kolodziej}
\affiliation{Advanced Photon Source, Argonne National Laboratory,  Argonne, Illinois 60439, USA}
\affiliation{National Synchrotron Radiation Centre SOLARIS, 30-392 Krakow, Poland}

\begin{abstract} 
Next-generation, high-brilliance x-ray photon sources call for new
x-ray optics. Here we demonstrate the feasibility of using monolithic
diamond channel-cut crystals as high-heat-load, beam-multiplexing,
narrow-band, mechanically-stable x-ray monochromators with high-power
x-ray beams at cutting-edge, high-repetition-rate x-ray free-electron
laser (XFEL) facilities. The diamond channel-cut crystals fabricated
and characterized in these studies are designed as two-bounce Bragg
reflection monochromators directing 14.4-keV or 12.4-keV x-rays within
a 15-meV-bandwidth to $^{57}$Fe or $^{45}$Sc nuclear resonant
scattering experiments, respectively. The crystal design allows
out-of-band x-rays within a $\simeq 1$-eV XFEL bandwidth to be
transmitted with minimal losses to alternative simultaneous
experiments. Only $\lesssim 2$\% of the incident $\simeq 100$-W x-ray
beam is absorbed in a 50-$\mu$m-thick first diamond crystal reflector,
ensuring that the monochromator crystal is highly stable. Other
x-ray optics applications of diamond channel-cut crystals are
anticipated.
\end{abstract}

\pacs{41.50.+h,41.60.Cr, 61.05.cp}
%
%X-ray optics, 41.50.+h
%Free-electron lasers, 41.60.Cr
%Diffraction -x~ray, 61.05.cp 
%X-ray lasers, 42.55.Vc
%

\maketitle

\section{Introduction}

Hard x-ray free-electron lasers (XFELs) in the self-seeding mode
\cite{SSSY,GKS11} generate fully coherent brilliant x-ray beams with a
well-defined and narrow spectrum, typically with a bandwidth of
$\simeq 0.2-1$~eV \cite{HXRSS12,MNY19,IOH19,NMO21} and a pulse
duration of a few tens of femtoseconds.

When operated in a high-repetition-rate ($\simeq$ MHz) pulse sequence
mode with properly tapered undulators, self-seeded XFELs may
deliver x-rays with a pulse energy of $\simeq 10$~mJ/pulse ($\simeq
7\times 10^{12}$~photons/pulse of $\simeq 10$-keV photons) and a
time-averaged power of more than 100~W, corresponding to a time-averaged
spectral flux of $\simeq 10^{17}$~photons/s in a bandwidth of
$\lesssim 1$~eV \cite{CGK16}.  This flux is about three orders of magnitude
larger than the average spectral flux currently possible with
storage ring--based synchrotron radiation sources [see also
  \cite{YS13,GKS15}].

An x-ray source with an increase of three orders of magnitude in the
spectral flux opens exciting new opportunities for hard x-ray
spectroscopic techniques, such as inelastic x-ray scattering \cite{Baron16}, x-ray photon correlation spectroscopy \cite{Shpyrko14} and 
nuclear resonant scattering \cite{RRAC16} in particular.

Such opportunities can be realized today, because high-repetition-rate
hard x-ray XFELs are now a reality.  The European XFEL in Hamburg
(Germany) \cite{EXFEL20} is the first operational high-repetition-rate
XFEL facility. The LCLS-II-HE in Stanford (USA) \cite{Raubenheimer18}
is coming soon.

However, these new hard x-ray sources with high average spectral
brilliance present not only opportunities but also challenges. One
challenge is how to monochromatize XFEL beams of very high average and
peak power.

To address these challenges, we have designed, manufactured and tested
diamond channel-cut crystals for use as high-heat-load,
beam-multiplexing, narrow-band (15-meV-bandwidth), mechanically-stable
monochromators.

\begin{figure}
\includegraphics[width=0.5\textwidth]{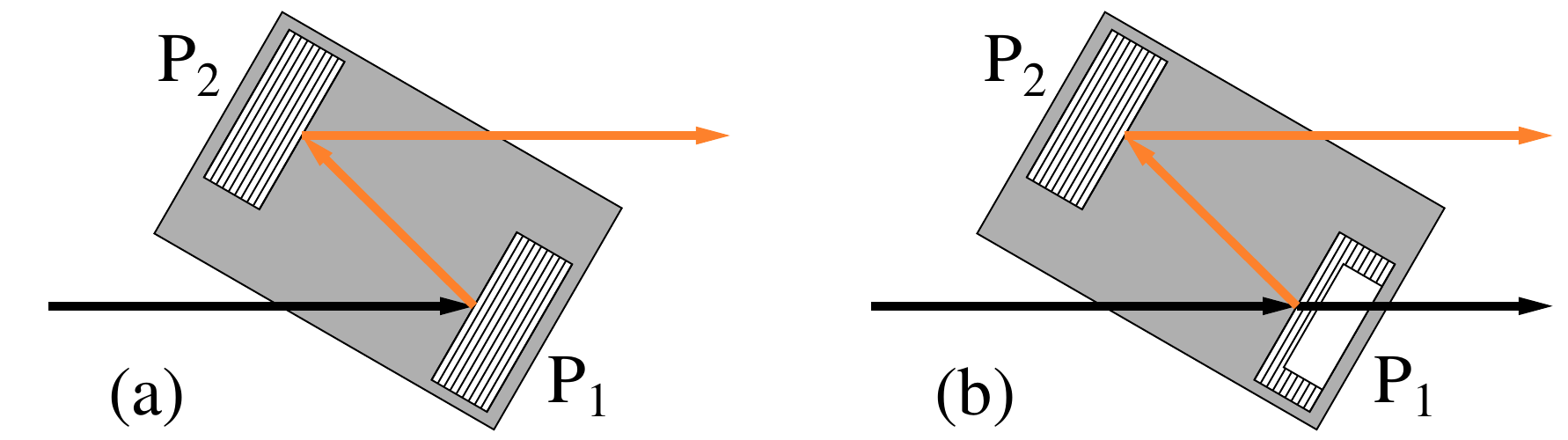}  
\caption{Schematic of monolithic channel-cut crystals with two
  reflecting crystal plates P$_1$ and P$_2$. Parallel lines in P$_1$
  and P$_2$ indicate Bragg-reflecting atomic planes. (a) Standard
  two-plate channel-cut configuration.  (b) design discussed here,
  featuring a thin crystal membrane in a drumhead structure on plate
  P$_1$. The membrane thickness is sufficient for high Bragg
  reflectivity in a narrow Bragg-reflection bandwidth and high
  transmissivity of the out-of-band x-rays. }
\label{fig1}
\end{figure}

A schematic of our channel-cut crystal design together with relevant
x-ray beams is shown in Fig.~\ref{fig1}(a).  We choose a channel-cut
monolithic monocrystalline design for two reasons.  First, this design
enables multiple successive Bragg reflections (two, four, etc.) by
perfectly aligned, parallel, mechanically stable reflecting atomic
planes in the reflecting crystal plates P$_1$ and P$_2$ of the
monolithic system. Second, it ensures parallel propagation of the
incident and multiply reflected beams.  The design includes a drumhead
structure shown in Fig.~\ref{fig1}(b) to permit one crystal face to be
an ultrathin membrane, as discussed later.

We choose diamond because diamond crystals have a unique combination of
outstanding physical properties that are perfect for many x-ray optics
applications for which traditional materials, such as silicon, fail to
perform \cite{SBT17}. Single crystals of silicon are attractive
because they have an almost perfect crystal lattice, are commercially
available in large quantities and are commonly used in x-ray crystal
optics. However, many applications, including those considered here,
require better performance than silicon can deliver: specifically,
better transparency to x-rays, resilience to radiation damage,
mechanical stiffness, x-ray Bragg reflectivity and thermal
conductivity.  Diamond is a material of choice in such cases, as it is
superior to silicon because of its unrivaled radiation hardness, an
order of magnitude higher transparency to x-rays, an order of
magnitude higher stiffness \cite{Field93,PPB98}, orders of magnitude
higher thermal conductivity \cite{Wei93, ITR18}, small thermal
expansion \cite{SSh10,SSh11} and almost 100\% reflectivity in Bragg
diffraction even in backscattering \cite{SSC10,SSB11}. This unique
combination of outstanding properties makes diamond the most promising
material for the transparent, resilient, high-resolution,
wavefront-preserving x-ray optics components that are essential for
many applications at third-generation light sources. These
characteristics are especially important at next-generation
high-brilliance, high-coherence light sources, such as
diffraction-limited storage rings (DLSRs) and XFELs.

The paper is organized as follows. Section~\ref{function} covers the function and optical design of the 
diamond channel-cut crystals. Details of fabrication are
presented in Section~\ref{manufacturing}. Results of characterization
of the crystals by x-ray rocking curve imaging are presented in
Section~\ref{rci}, and results of x-ray performance tests are given in Section~\ref{performance}.
Conclusions and outlook are discussed in Section~\ref{conclusions}.

\section{Function and optical design}
\label{function}

High-repetition-rate (HRR) hard x-ray self-seeded (HXRSS) XFELs will
provide x-ray photon beams with an unprecedented spectral flux density
advantageous for many applications, in particular for high-resolution
spectroscopies, such as nuclear resonant scattering (NRS), inelastic x-ray
scattering (IXS) and others. However, dealing with an x-ray beam of such
high power is a challenge. Diamond channel-cut crystals can help
solve this problem. Our recipe for the solution has four ingredients: a high-indexed
Bragg reflection (for a narrow-band reflection), diamond crystal (for
x-ray transparency and excellent thermomechanical properties),
a drumhead design (for small absorbed power and beam sharing) and
a monolithic channel-cut design (for in-line geometry and stability).

An important attribute of high-resolution spectroscopies is the use of
x-rays beams with milli-electron volt or even sub-milli-electron volt
bandwidths. As a result, such techniques require x-ray sources with
large spectral density, and
ultra-high-resolution x-ray crystal monochromators with a bandpass of
1\textendash 0.1 meV along with submicroradian angular stability; see
\cite{Shvydko-SB} for a review.  Such optics usually cannot perform
under high-heat-load conditions because of the resulting substantial
angular instabilities. At storage ring synchrotron radiation
facilities, additional primary or high-heat-load crystal
monochromators are used \cite{CSC14}. This approach reduces the
bandwidth of the undulator x-rays to about 1\textendash2~eV and thus
reduces substantially (by a factor of $\simeq 100$) the heat load on
the ultra-high-resolution x-ray crystal monochromators.

However, the standard high-heat-load monochromators
are not helpful to deal with x-ray beams generated by HRR-HXRSS
XFELs, as these sources typically already have a relatively narrow bandwidth of
$\simeq 1$ eV. Rather, such beams require a high-resolution
monochromator with a bandpass of $\simeq 10$~meV that is stable under high heat loads. This smaller bandpass would reduce the heat load on the ultra-high-resolution optic
by a factor of about 100.  As an example, 10-keV x-rays can be
monochromatized to $\Delta E \simeq 10$~meV with a single Bragg
reflection\footnote{XFEL  beams typically have a small angular
  divergence of $\Delta\theta \simeq 2~\mu$rad
  \cite{SSY,HK07}. Because of this property, monochromatization of x-rays to a
  $\simeq 10$~meV bandwidth can be achieved with a single high-indexed
  Bragg reflection, because the spectral broadening due to the angular
  divergence can be kept small ($\delta E/E \simeq
  \Delta\theta/\tan\theta \lesssim 10^{-6}$), provided Bragg's angle
  $\theta$ is large and $\tan\theta \gg 1$. The use of high-indexed Bragg
  reflections with $\theta$ close to 90$^{\circ}$ is therefore
  important.}. In this case, about 1\% of the XFEL radiation
power in the small bandwidth can be directed to the
ultra-high-resolution monochromator and associated spectroscopic
experiments. The remaining 99\% of the incident flux is transmitted, rather than being deposited in the crystal, thus avoiding crystal heating, the
resulting angular instabilities and, therefore, angular instabilities
in the reflected beam. If the monochromator is suitably designed, this remaining flux can be directed to other experiments.

Such multiplexing of the beam requires an x-ray-transparent
Bragg-reflecting crystal. Such a crystal would allow most of the
incident x-rays, with the exception of the Bragg reflections in the
narrow bandwidth, to pass through without absorption. Silicon cannot
perform in this way because the 10 extinction lengths required for
maximum Bragg reflectivity are comparable to the photoabsorption
length in silicon. In contrast, for diamond crystals, 10 extinction
lengths equal about 1/10th of the photoabsorption length. Exactly
because in diamond the extinction length is much smaller than the
absorption length, diamond crystals experience close to 100\% Bragg
reflectivity, even in backscattering, and simultaneously offer high
x-ray transparency \cite{SSB11}.  Therefore, we chose to fabricate a
diamond crystal membrane with a thickness of about 5 extinction
lengths for high x-ray reflectivity and transparency.

However, this dimension of 5 extinction lengths measures only about 20
to 100~$\mu$m, depending on the chosen Bragg reflection. It is
challenging to fabricate and handle such ultrathin crystal components
without introducing crystal damage and strain. A solution is to use a
drumhead crystal: a monolithic crystal structure comprised of a thin
membrane furnished with a surrounding solid collar, shown
schematically in Fig.~\ref{fig1}(b). This design ensures mechanically
stable, strain-free mounting of the membrane while still allowing
efficient thermal transport \cite{KVT16}.

Finally, to ensure a fixed direction of the reflected x-rays
independent of Bragg's angle, we add one more equivalent Bragg
reflection from another crystal plate in a non-dispersive setting, as
shown in Fig.~\ref{fig1}. To ensure mechanical stability, both crystal
plates are a part of a single, monolithic, channel-cut crystal
structure.  Channel-cut crystals have been popular in x-ray optics
applications since the 1960s
\cite{BH65,BH65-2,BH66,Deslattes68}. Typically they are fabricated
from silicon or germanium crystals.  Manufacturing a channel-cut
crystal from diamond is a challenge, but becomes possible using
advanced modern techniques, as discussed in
Section~\ref{manufacturing}.

\begin{figure}
  \includegraphics[width=0.5\textwidth]{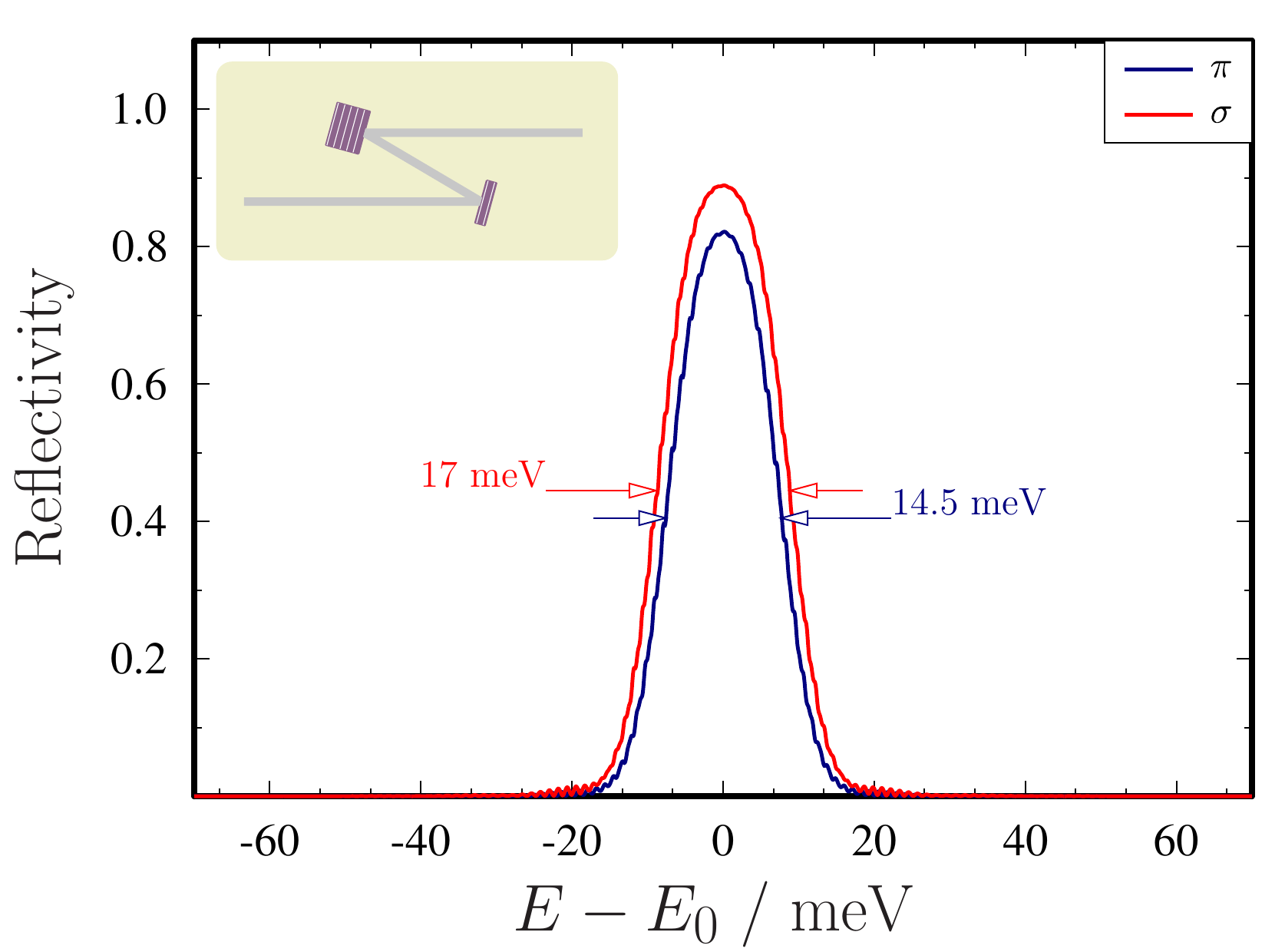}
\caption{Calculated x-ray photon energy dependencies for two successive 800 Bragg reflections from a diamond channel-cut crystal. The inset shows the optical setup. The central energy, $E_{\ind{0}}=14.4125$~keV, corresponds to the nuclear resonance of $^{57}$Fe. Bragg's angle is $\theta=74.7207^{\circ}$. The spectral dependences are calculated for two possible x-ray polarization states $\pi$ and $\sigma$}
\label{fig110}
\end{figure}

In the present discussion, we focus on a diamond monolithic
channel-cut crystal for use with 14.4125~keV x-rays for $^{57}$Fe
nuclear resonant experiments \cite{GdW,RR04,RRAC16}. We also explored
a similar design for use with 12.4~keV x-rays for $^{45}$Sc experiments \cite{SS90}.

The 733 Bragg reflection in diamond is the highest indexed reflection that is 
accessible with 14.4-keV x-rays; it has the
largest Bragg's angle ($\theta=80.75226^{\circ}$) and the smallest Bragg
reflection bandwidth (13~meV). However, because of practical considerations
related to channel-cut manufacturing (discussed in
Section~\ref{manufacturing}), we opted for the reflection with the next highest index, the 800 Bragg reflection,
with a smaller Bragg's angle of $\theta=74.7207^{\circ}$ and a larger bandwidth of 20~meV.

Figure~\ref{fig110} shows predicted x-ray photon energy dependencies
of the double 800 Bragg reflection profiles from a diamond channel-cut
crystal, numerically calculated using equations of dynamical theory of
x-ray diffraction in crystals \cite{Authier,Shvydko-SB}. The first and
second reflections are from 50-$\mu$m- and 500-$\mu$m-thick crystal
plates, respectively. The energy bandwidth $\Delta E$ is 17~meV for
the $\sigma$-polarization and 14.5~meV for the $\pi$-polarization
components of the beam. An angular divergence of the incident x-rays
of $2.5~\mu$rad is used in the calculations. The calculations show
that only 2\% of the incident x-ray beam is absorbed in the first
plate (the 50-$\mu$m-thick diamond drumhead crystal), thus making
possible high stability of the monochromator crystal.

Such a diamond monolithic channel-cut crystal can function as a
two-bounce 800 Bragg reflection monochromator, deflecting 14.4~keV
x-rays within a $\simeq 15$-meV bandwidth to $^{57}$Fe nuclear
resonant scattering experiments, while the rest of the x-rays in the
1-eV XFEL bandwidth are transmitted through the thin drumhead crystal
and could be transported to a simultaneous experiment \footnote{To
  ensure further stability of such a multiplexed setup, it would be
  advantageous to choose the scattering plane for the diamond
  channel-cut crystal to be perpendicular to the scattering plane of
  the downstream ultra-high-resolution optic, as discussed in
  \cite{CGK16}.}.

\begin{figure}
\includegraphics[width=0.5\textwidth]{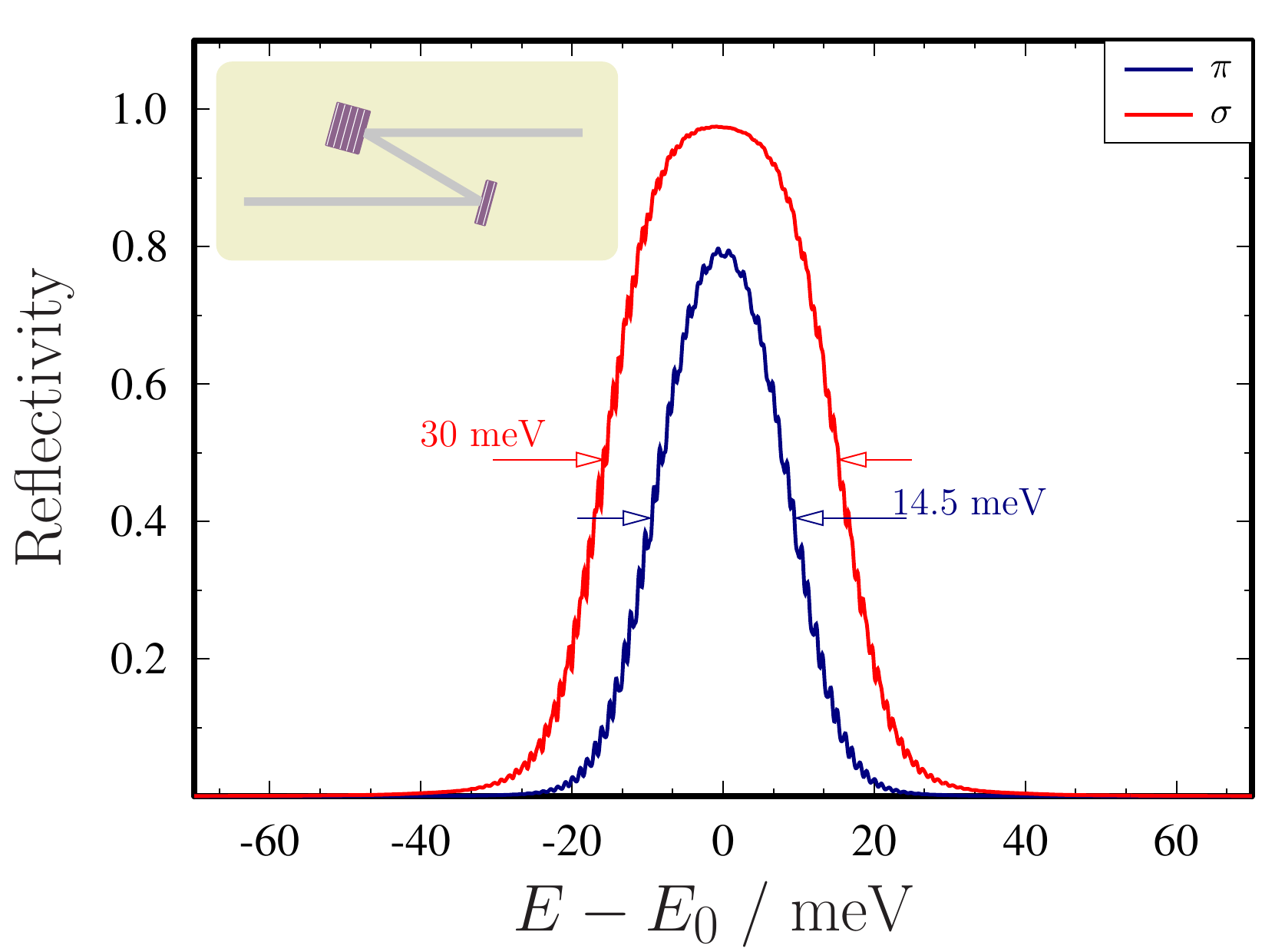}  
\caption{Calculated x-ray photon energy dependencies for two successive 620 Bragg reflections from a diamond channel-cut crystal. The inset shows the optical setup. The central energy, $E_{\ind{0}}=12.4$~keV, corresponds to the nuclear resonance in $^{45}$Sc. Bragg's angle is $62.424^{\circ}$.}
\label{fig115}
\end{figure}

Similarly, 620 Bragg reflections are suitable for a diamond channel-cut
crystal narrow-band monochromator for 12.4~keV x-rays, corresponding
to the nuclear resonance in $^{45}$Sc \cite{SS90}; see
Fig.~\ref{fig115}.

\section{Manufacturing}
\label{manufacturing}

Two conditions are key to successful
realization of a properly functioning diamond channel-cut crystal
x-ray monochromator: (1) the availability of flawless, single-crystal diamond material in
sufficiently large workable sizes and (2) the ability to machine diamond
without introducing crystal defects. 

\begin{figure}%
  \includegraphics[width=0.5\textwidth]{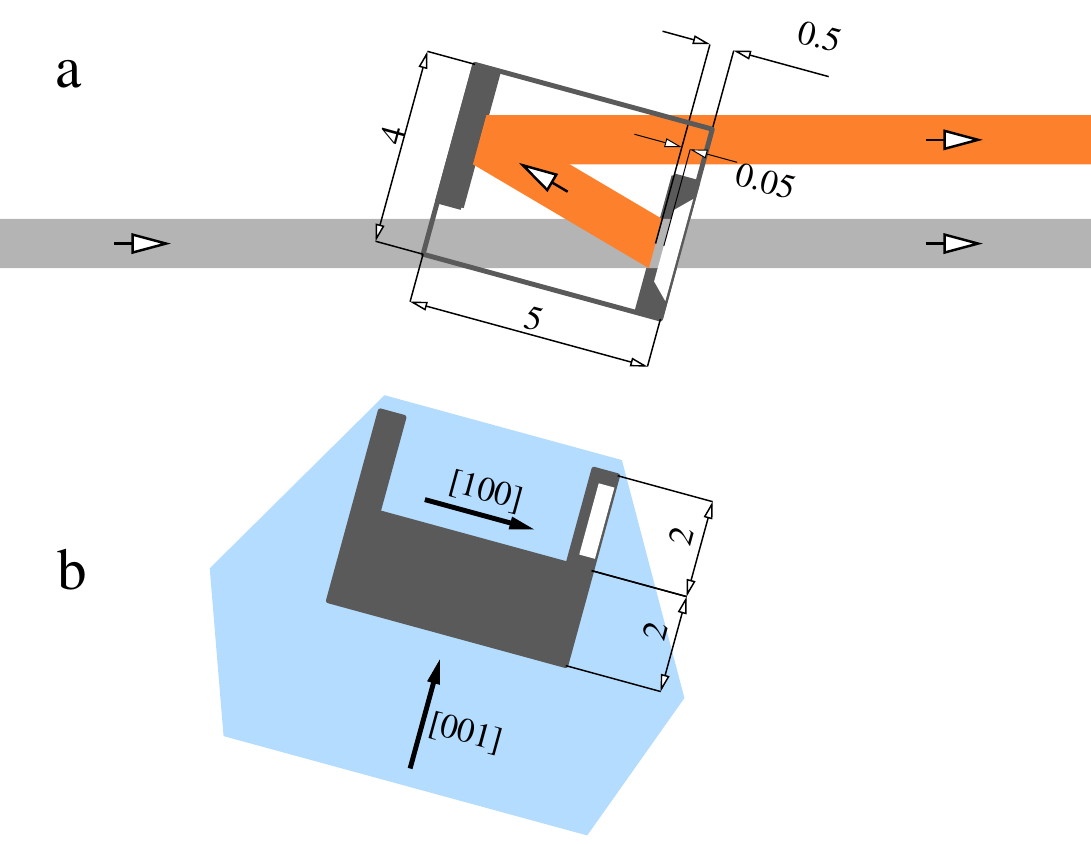} 
\caption{Schematic of the (800) diamond channel-cut crystal
  monochromator for 14.4-keV x-rays (black). The channel cut is machined from an as-grown
  cubo-octahedral diamond crystal stone (light blue).  The
  graph shows top (a) and side (b) views with dimensions in
  millimeters. Propagation paths of x-ray beams with a 1-mm
  cross-section are shown in (a).}
\label{fig120}
\end{figure}

The characteristic linear dimensions of  presently available 
high-quality synthetic diamond crystals are in the range of a few
millimeters [see \cite{SBT17} for a recent review]. The crystals are
relatively small but are sufficient for use with XFEL beams,
which are typically less than 1~mm in cross-section.
Figure~\ref{fig120} shows a schematic of the diamond
channel-cut crystals produced in the present studies.

The channel-cut crystals were cut from synthetically grown raw diamond
stones of type IIa (with nitrogen content below 1~ppm).  The diamond
stones were grown using a temperature gradient method under high
pressure and high temperature (HPHT) conditions.  The stones used in these
studies were grown and machined at the
Technological Institute for Superhard and Novel Carbon Materials
(TISNCM) in Troitsk, Russia \cite{BKN07,PDK11}.

Raw HPHT-grown diamond stones typically have a cubo-octahedral
form. It is well known that the top part of the (001) cubic growth
sector (the area furthest from the 001 crystal seed) features the
highest crystal quality with lowest defect density
\cite{BCC09,ST12}. Therefore, in these studies, the channel-cut
crystals are extracted from the top part of the stone, as indicated in
Fig.~\ref{fig120}(b). With this cut, the crystal plate surfaces
parallel to the atomic planes (hk0) and to the [001] direction take
the largest volume of highest quality. For this reason, we chose to
make a (800) monochromator for use with 14.4~keV x-rays, instead of
the narrower band (733) design. Additional reasons for the (800)
choice are related to laser machining, as discussed later. Similarly,
we chose to make a (620) channel-cut monochromator for 12.4~keV x-rays
in the $^{45}$Sc case.

Production of the monochromator crystals begins with selection of
high-quality, as-grown diamond crystal stones for machining by visual
screening. To ensure the fewest defects, crystals with well-developed,
defect-free cubic facets are selected, because, as noted previously,
the (100) cubic growth sector contains the smallest amount of
contamination and structural defects.  The diamond stones are machined
with the newest laser machining technologies.

The monochromator atomic planes [(800) or (620)] of the selected
diamond stones are oriented to an accuracy of better than
0.5$^{\circ}$, and the channel-cut crystals are cut to the designed
shape using a pulsed Nd:YAG laser with the following parameters: 532
nm wavelength (second harmonic), 15~W average power, 10 kHz repetition
rate, $\simeq$~1.5~mJ pulse energy, 40 ns pulse duration and 15~$\mu$m
focused spot size (full width half maximum [FWHM]).  The laser beam is
positioned and steered on the workpiece with a two-coordinate optical
mirror scanning system equipped with an F-Theta scanning lens. The
laser beam is moved with a linear speed of 100~mm/s.

\begin{figure}
  \includegraphics[width=0.5\textwidth]{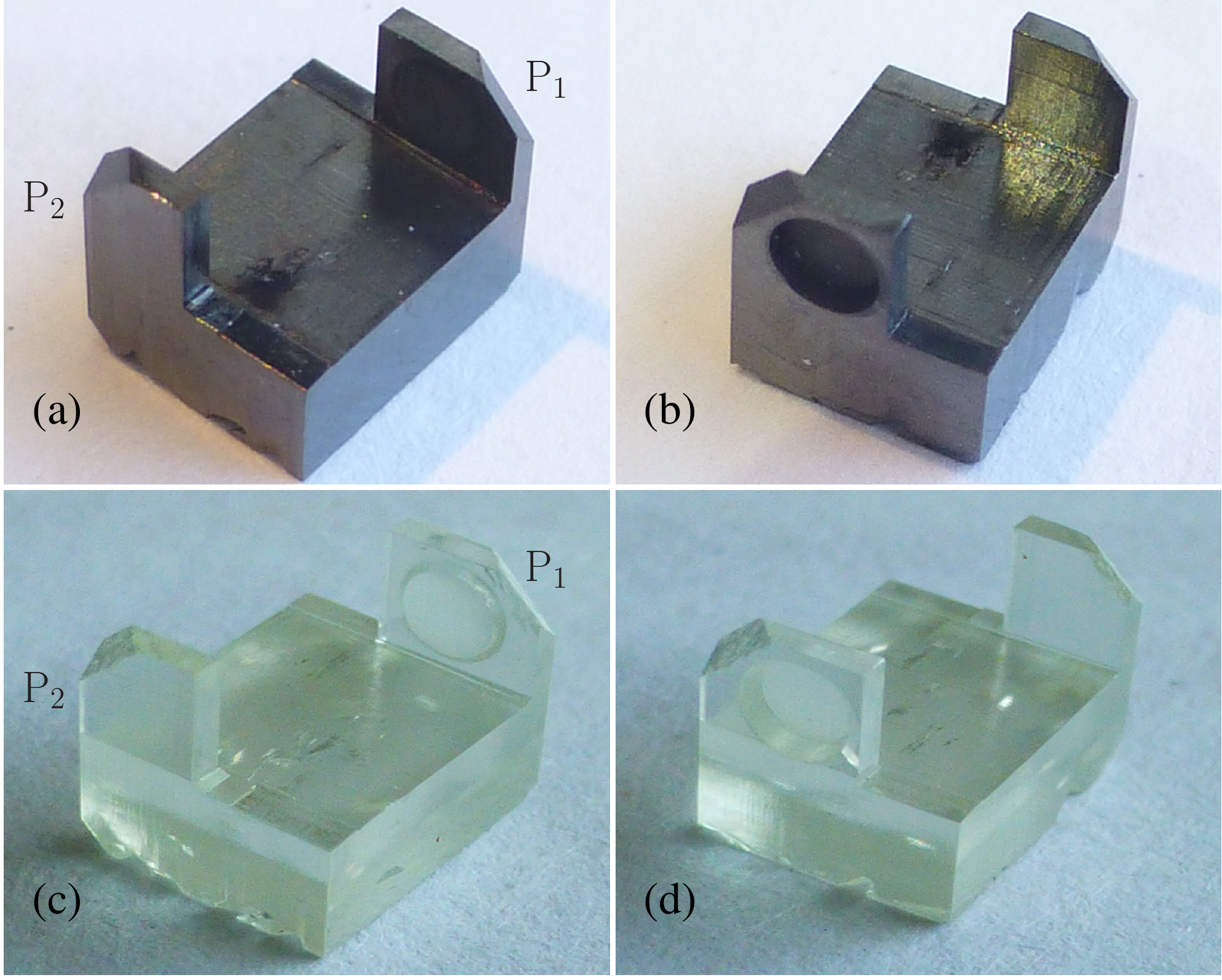}
\caption{Photographs of the first (800) channel-cut crystal taken after
  laser machining (a)-(b) and after annealing in air at 630$^{\circ}$C for 3
  hours (c)-(d). }
\label{fig2}
\end{figure}

The drumhead structure is formed by layer-by-layer ablation with
picosecond pulses from a Nd:YAG laser operated in the third harmonic
(355 nm wavelength) with the following parameters: 1~W average power,
10~ps pulse duration, 500~kHz repetition rate and 2~$\mu$J pulse
energy.  The use of the third harmonic enables reduction of the
focused beam spot size and most importantly reduction of the diameter
of a crater in diamond produced by a single laser pulse.  The crater
diameter is 8~$\mu$m, provided the beam is focused on the diamond
surface.  During ablation, the beam is steered over the workpiece at a
speed of 150~mm/s with a similar two-coordinate scanning system with
an F-Theta lens. The removal rate is about 0.3~$\mu$m per layer. The
surface roughness of the drumhead surface after laser ablation of a
0.5-mm-thick crystal is $\lesssim 1~\mu$m (rms). Similar laser
parameters and procedures have been used previously to manufacture
diamond x-ray lenses \cite{TBP15} and drumhead crystals \cite{KVT16}.

Figure~\ref{fig2}(a) shows photographs of the first machined (800)
channel-cut crystal. The crystals appear black
after machining. Apparently, laser machining is accompanied by
graphitization of the diamond surface. To erase the blackening, we
anneal the crystals in air at a temperature of 630--650$^{\circ}$C
for 3 hours, as in our previous and recent studies \cite{KVT16,PWH20}.
After annealing, the blackening has vanished, as seen on the photographs
in Figs.~\ref{fig2}(c)-(d).  Most important, annealing erases any
crystal strain induced in the process of laser cutting or ablation,
which is critical for the proper functioning of the channel-cut
crystal as a narrow-band x-ray monochromator. The annealing
temperature is chosen such that all residuals of graphite and other
carbon compounds are burned in air, while keeping diamond
intact\footnote{Increasing annealing temperature or time may result in
  etching of the crystal surface.}.  We will refer to this procedure
as medium-temperature in-air annealing (MTA).

The first reflecting crystal plate of the channel-cut crystal (P$_1$ in
Fig.~\ref{fig2}) features a 50-$\mu$m-thick drumhead, designed to
achieve high reflectivity ($\gtrsim 90\%$) of x-rays in the
Bragg-reflection band and 98\% transmissivity of the out-of-band
x-rays. X-rays in the 15-meV band reflected from both crystal plates
(P$_1$ and P$_2$) of the channel-cut crystal (see Fig.~\ref{fig110})
are directed to a high-spectral-resolution x-ray instrument.  The
major part, 98\% of the incident x-rays in a 1-eV bandwidth
\cite{CGK16,GKS15}, is transmitted and can be used in
simultaneous parallel experiments. The 2-mm-thick base
allows the crystal to be mounted mechanically tight and
strain-free and to be water-cooled efficiently, ensuring stable operation in high-power XFEL beams.

\begin{figure}
\includegraphics[width=0.5\textwidth]{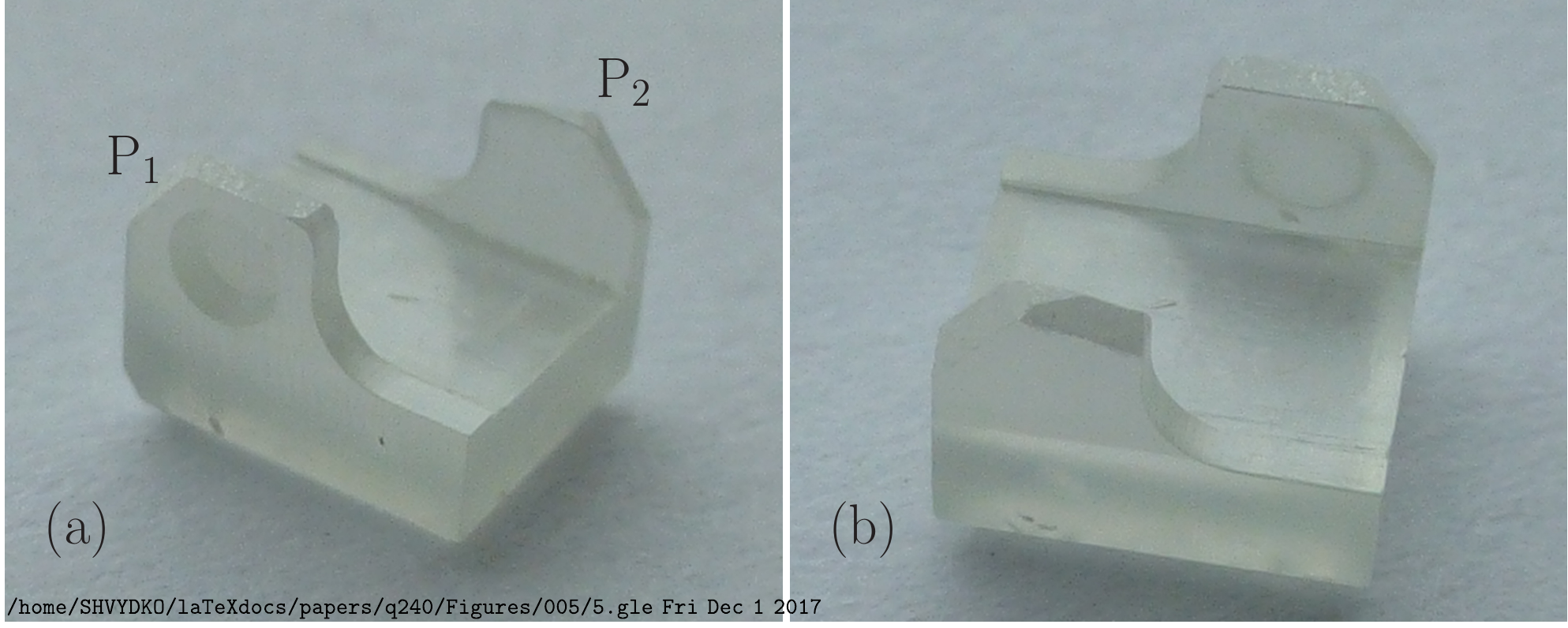}
  \caption{Photographs of the (620) channel-cut crystal after annealing.
}
\label{fig8}
\end{figure}

Figure~\ref{fig8} shows an example of a (620) diamond channel-cut
crystal designed and manufactured for $^{45}$Sc nuclear resonant
experiments with 12.4~keV x-rays. The (620) channel-cut crystal also
features a 50-$\mu$m drumhead on the first plate P$_1$.  Making
drumhead crystals in orientations different from (100) is a challenge
\cite{KVT16}.

Anisotropy of the physical properties of diamond crystals leads to
different results in laser ablation of surfaces with different crystal
orientations. Surfaces with the (100) orientation do not present
significant problems for laser ablation. In contrast, surfaces with
the (111) crystallographic orientation are most difficult both for
laser ablation and for polishing. Laser ablation of the (111) surface
creates microcracks.

The quality of the (620) surfaces after laser ablation is typically
worse than that of the (100) surfaces. We found that increasing the
crystal temperature improves the surface quality of the machined
surfaces. During machining of the drumhead with the (620) surface
orientation, the crystal is heated to 600$^{\circ}$C by a resistive
heater installed in the crystal holder.

\begin{figure}
\includegraphics[width=0.5\textwidth]{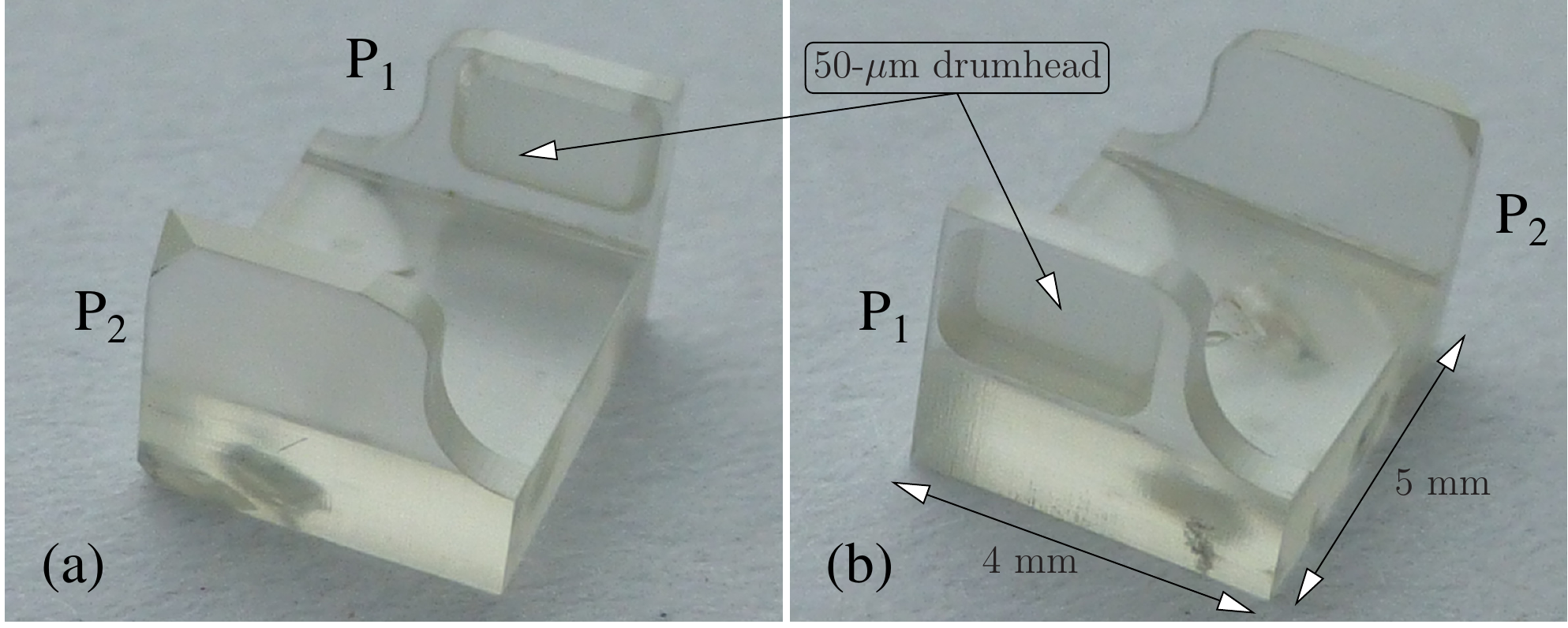}  
\caption{Photographs of the best-quality (800) channel-cut monochromator crystal (third iteration) after annealing. Note the optimized rectangular shape of the drumhead area. 
  %[C001-180202]
}
\label{fig6}
\end{figure}

Altogether, three (800) and two (620) channel-cut monochromator
crystals were manufactured and characterized. In the following
section, we present x-ray characterization results of the best (800)
channel-cut crystal monochromator (third iteration), shown in Fig.~\ref{fig6}

\section{Characterization of channel-cut crystals}
\subsection{Crystal quality test: X-ray rocking-curve imaging}
\label{rci}

\begin{figure}
  \includegraphics[width=0.5\textwidth]{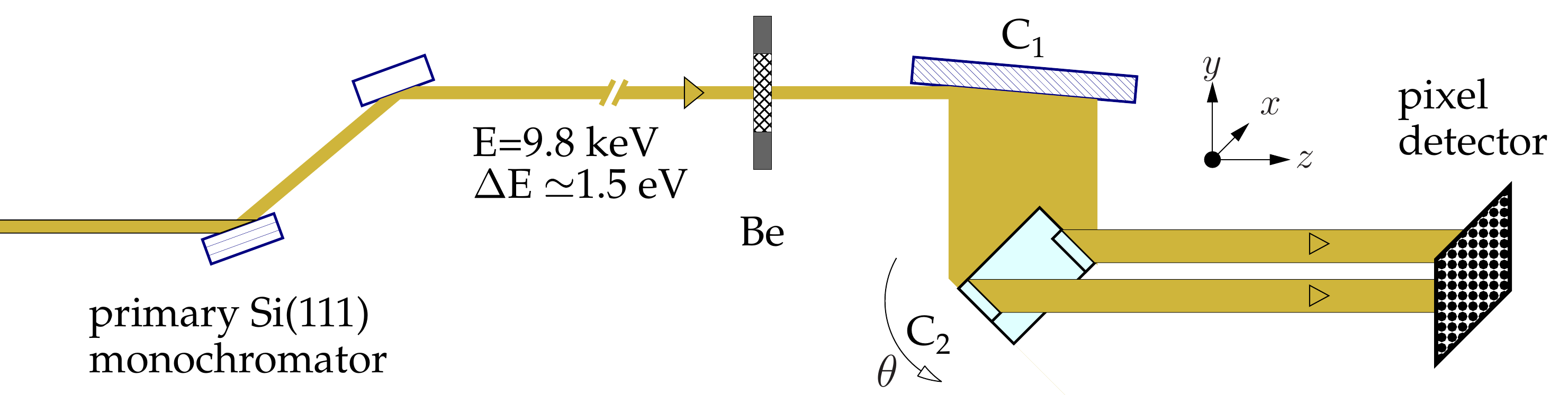}  
\caption{The layout and optical components of the rocking curve
  imaging setup at APS bending magnet
  beamline 1BM, comprising a primary double-crystal Si(111)
  monochromator, beryllium (Be) window, Si conditioning crystal
  C$_{\ind{1}}$ in the asymmetric 531 reflection, diamond (800)
  channel-crystal under study C$_{\ind{2}}$ in the 400 Bragg
  reflection and pixel detector. X-rays with a 9.8-keV photon energy
  and a 1.5-eV bandwidth are used.}
\label{fig9}
\end{figure}

The quality of the diamond channel-cut crystals was characterized
using x-ray Bragg diffraction rocking curve imaging (RCI), also known
as sequential topography \cite{LBH99}.  In this technique, Bragg
reflection images of a crystal under study are measured with a pixel
x-ray detector, with images being taken sequentially at different
incidence angles to the reflecting atomic planes of the
well-collimated x-rays; Bragg reflection maps are calculated for the
reflection peak intensities, angular widths and angular positions. We
used an RCI setup \cite{SST16,PWH20} at x-ray optics testing beamline
1BM \cite{Macrander2016} at the Advanced Photon Source (APS),
schematically shown in Fig.~\ref{fig9}.

The RCI setup employs a nearly non-dispersive double-crystal
C$_1$-C$_2$ arrangement. The first conditioning crystal C$_1$ is an asymmetrically cut, high-quality silicon crystal in the 531
Bragg reflection; the second crystal C$_2$ is the structure under study, 
the (800) diamond channel-cut crystal, in the 400
Bragg reflection. We used the 400 Bragg reflection rather than the 800
reflection to characterize the (800) channel-cut crystal for the
following reason. We want both reflecting plates
P$_1$ and P$_2$ of the channel-cut crystal to be illuminated and
imaged simultaneously so that we can evaluate whether there is any angular
misorientation of the plates due to crystal defects.  To meet this requirement, the Bragg angle
must be chosen close to  $45^{\circ}$, which in turn would require
high-energy (19.66-keV) x-ray photons if the 800 reflection were
used. Such a choice would result in time-consuming measurements, as
the photon flux is relatively low at this high photon energy, and the
angular reflection width ($\Delta\theta_{\ind{800}}=1.2~\mu$rad) is very
narrow (compare to $\Delta\theta_{\ind{800}}=4.7~\mu$rad for 14.4-keV
x-rays).  Therefore, we chose instead the 400 Bragg reflection and
9.831-keV x-rays, for which Bragg's angle
$\theta_{\ind{400}}=45^{\circ}$, with angular width
$\Delta\theta_{\ind{400}}=9.1~\mu$rad if the crystal is 0.5 mm thick (reflecting plate P$_1$),
and $\Delta\theta_{\ind{400}}=9.8~\mu$rad if the crystal is 0.05 mm
thick (drumhead area of reflecting plate P$_2$).

\begin{figure}
  \includegraphics[width=0.5\textwidth]{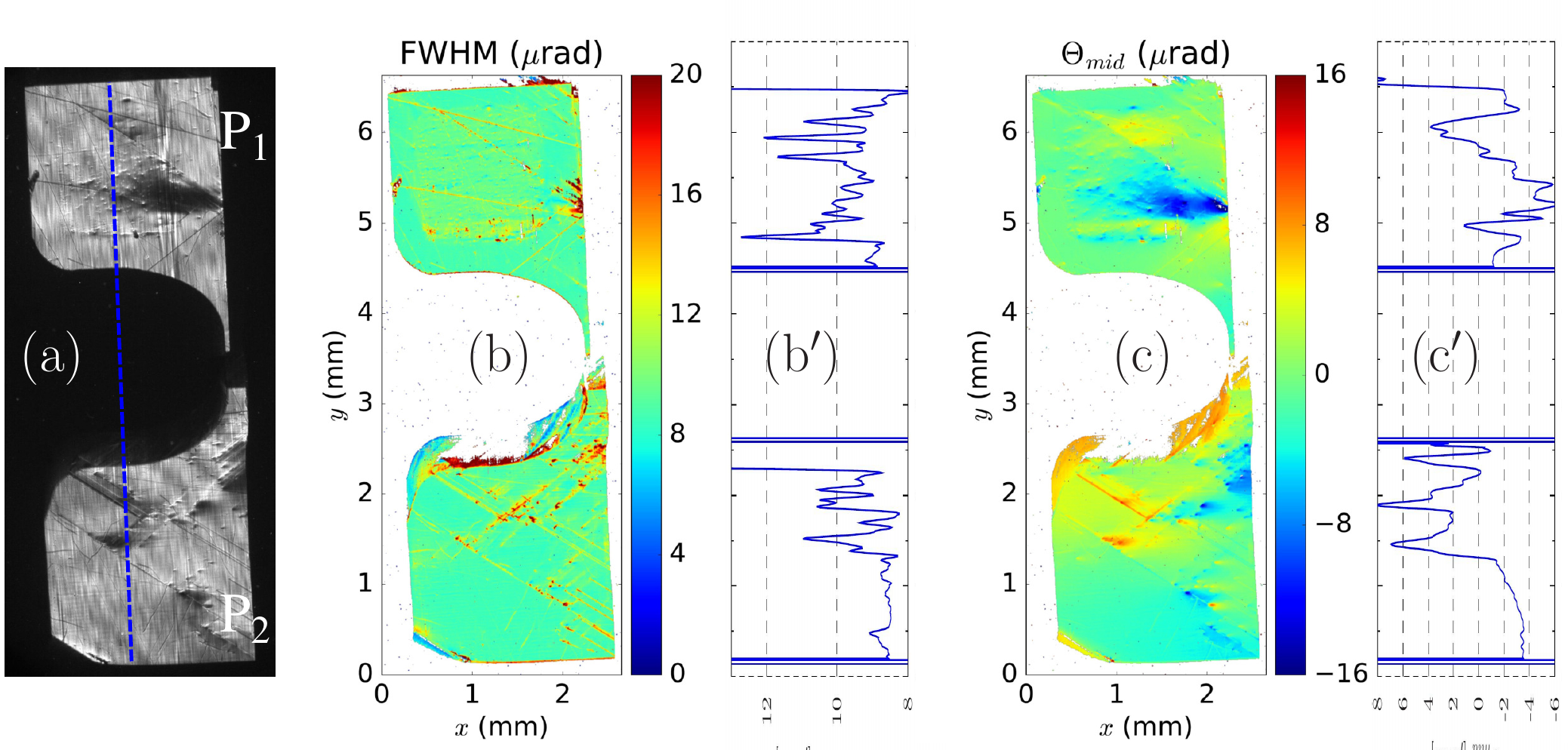}
  \caption{The 400 Bragg reflection x-ray topography and the RCI maps of the diamond (800) channel-cut crystal measured after
    second MTA annealing.  X-rays illuminate the front face of plate P$_1$
    and the rear face of plate P$_2$ (drumhead), respectively, corresponding to
    Fig.~\ref{fig6}(a).  (a) X-ray topography taken at the peak of the
    crystal-integrated rocking curve. (b) and (c) Color maps of the
    Bragg reflection angular widths (FWHM) and mid-point of the Bragg
    reflection angular curves (rocking curves), respectively.  White
    pixels indicate that the appropriate values of FWHM and mid-point
    are out of the color scale ranges. (b$^{\prime}$) and
    (c$^{\prime}$) Cross-section profiles of color
    maps (b) and (c), respectively, measured along the blue line
    indicated in (a).}
\label{fig7a}
\end{figure}

\begin{figure}
  \includegraphics[width=0.5\textwidth]{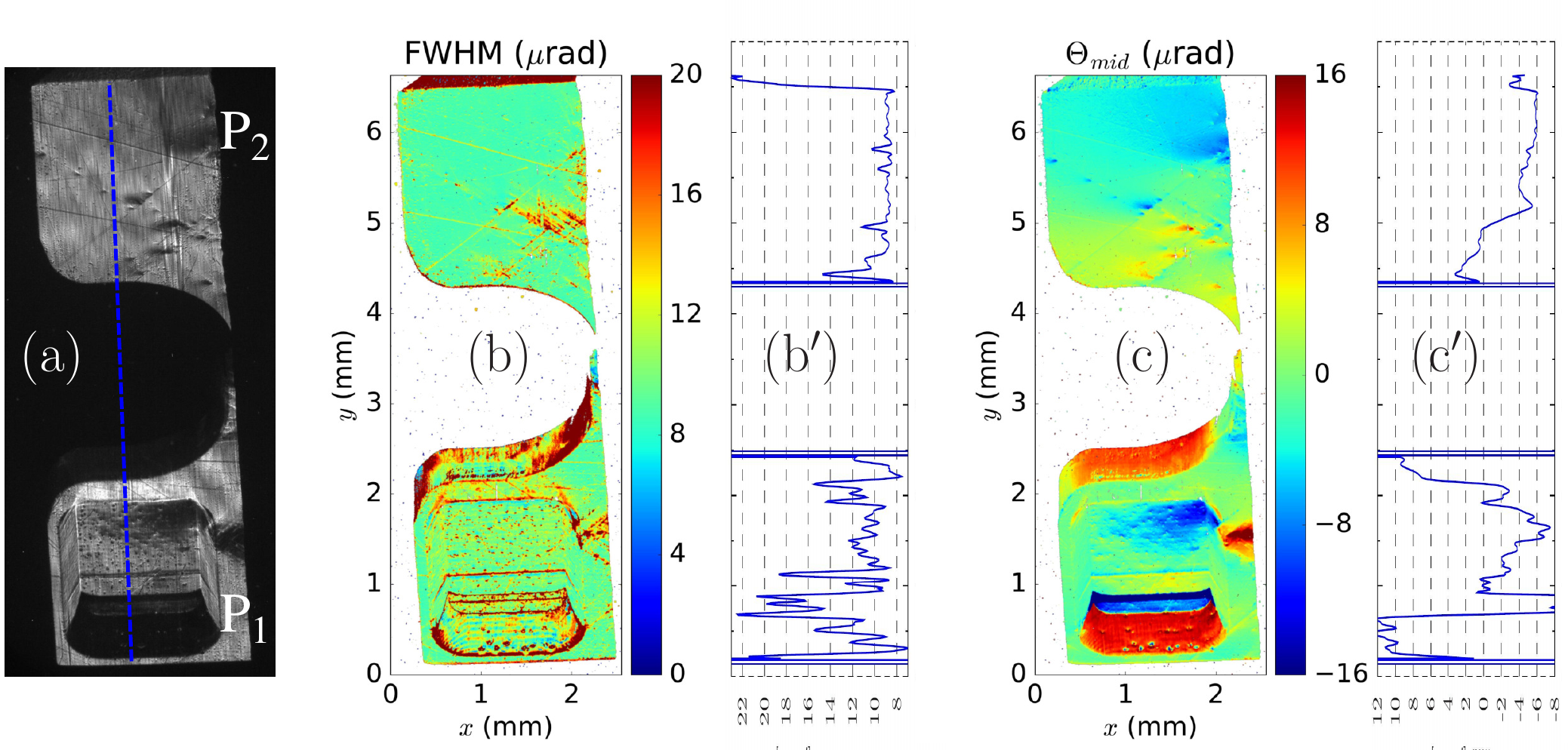}
\caption{Similar to Fig.~\ref{fig7a}, but with the channel-cut crystal exposed to x-rays from the reverse side, corresponding to Fig.~\ref{fig6}(b).}
\label{fig7b}
\end{figure}

The upstream double-crystal Si(111) monochromator is therefore
tuned to select 9.831-keV x-rays. The monochromator bandwidth is
$\simeq$~1.5~eV.  For the conditioning crystal C$_{\ind{1}}$, Bragg's
angle is $\theta_{\ind{531}}=43.385^{\circ}$. The asymmetry angle is
chosen as $\eta_{\ind{531}}=41.4^{\circ}$, resulting in the asymmetry parameter 
$b=-28.7$. This choice ensures $\simeq 1~\mu$rad angular collimation and
a x-ray beam size greater than 20 mm illuminating the
channel-cut crystal C$_{\ind{2}}$. Under these conditions, both plates of the
channel-cut crystal are illuminated and can be imaged simultaneously
in Bragg diffraction. The setup enables RCI mapping with 
submicroradian angular and 2.5-$\mu$m spatial resolution.

X-ray images of the (800) channel-cut crystal taken at the maximum of
the sample-integrated Bragg reflection curves are shown in
Figs.~\ref{fig7a}(a) and \ref{fig7b}(a). In Fig.~\ref{fig7a}, x-rays
illuminate the front (working) face of plate P$_1$ and the rear face
of plate P$_2$, respectively, as in Fig.~\ref{fig6}(a). In Fig.~\ref{fig7b}, the sample is exposed to x-rays from the reverse side, as in
Fig.~\ref{fig6}(b).

\begin{figure*}
\includegraphics[width=1.0\textwidth]{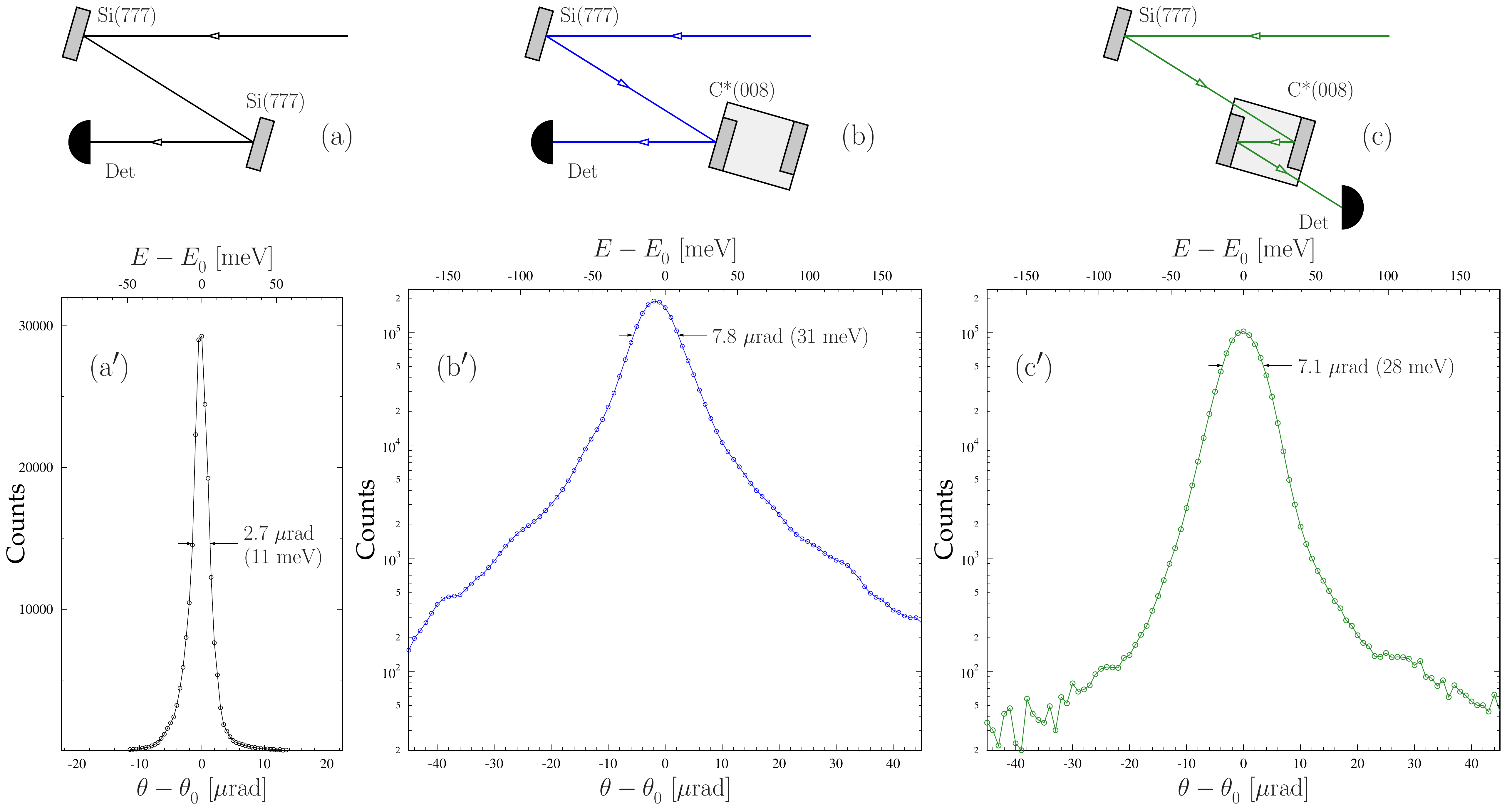}  
\caption{Schematics of multi-crystal Bragg-diffraction arrangements in
  non-dispersive settings (a), (b) and (c) to measure the angular
  dependences of Bragg reflection of 14.41-keV x-rays from
  (a$^{\prime}$) a Si(777) crystal, (b$^{\prime}$) from one (800)
  plate of the diamond channel-cut crystal, and (c$^{\prime}$) from
  both plates sequentially, respectively. In all cases x-rays are
  first reflected from a Si crystal in the (777) Bragg reflection in
  a non-dispersive setting with regard to the successive Bragg reflections.}
\label{fig11}
\end{figure*}

The angular dependencies of Bragg reflectivity (rocking curves)
measured with each detector pixel are used to calculate Bragg
reflection maps. Figures~\ref{fig7a}(b) and \ref{fig7b}(b) show color
maps of the angular widths (FWHM) of the
rocking curves.  The color maps of the rocking curves' mid-points
are shown in Figs.~\ref{fig7a}(b) and \ref{fig7b}(b). The maps are
calculated using a dedicated code \cite{Stoupin15}.  Cross-sections through the FWHM and mid-point maps along the blue lines in
Figs.~\ref{fig7a}(a) and \ref{fig7b}(a) are presented in
Figs.~\ref{fig7a}(c$^{\prime}$) and \ref{fig7b}(a), respectively.

The microscopic defects can be derived from the Bragg reflection FWHM
maps. The mesoscopic and macroscopic crystal strain and Bragg planes
slope errors can best be evaluated from the mid-point maps. In
particular, if the plates of the channel-cut crystal are misoriented
due to crystal defects, the mid-point maps will provide this
information, because both plates of the channel-cut crystal are imaged
simultaneously.

The x-ray images and RCI maps in Figs.~\ref{fig7a} and \ref{fig7b}
reveal good crystal quality of the manufactured (800) channel-cut
crystal, especially in its upper part. Nevertheless, some defects can
be detected. These defects result in increased angular widths and
varying angular position of the rocking curves. In the center of the
plates (in the working area), the angular reflection widths are
increased by a few microradians from the expected 9-$\mu$rad angular
width of the 400 Bragg reflection.  The angular position of the
rocking curves also varies within a range of a few microradians, which indicates
angular misalignment of the plates. These distortions are caused chiefly by the crystal defects in the as-grown crystal stone. Machining of the diamond has a secondary effect.

\subsection{X-ray performance test: Two-bounce Bragg reflection from the diamond channel-cut crystal}
\label{performance}

The performance of the manufactured (800) diamond channel-cut crystal
was tested by measuring the angular width $\Delta \theta$ of two
successive Bragg reflections of x-rays from its crystal plates
P$_{\ind{1}}$ and P$_{\ind{2}}$. The angular width is measured using a
non-dispersive double-crystal arrangement, with the first reference
crystal chosen to have an angular reflection width much smaller than
the angular width of the channel-cut crystal. The spectral width of
the channel-cut double reflection can be estimated as $\Delta E
/E=\Delta \theta/\tan\theta$ using the differential presentation of
Bragg's law and the measured $\Delta \theta$ value.

We used the 777 Bragg reflection of 14.412~keV x-rays from the reference
Si crystal with Bragg's angle $\theta_{\ind{777}}=73.7848^{\circ}$ and an
angular reflection width of $\Delta\theta_{\ind{777}}=1.2~\mu$rad.  It
matches well to the 800 Bragg reflection from diamond, with
Bragg's angle of $\theta_{\ind{800}}=74.7206^{\circ}$ and an angular width
of $\Delta\theta_{\ind{800}}=5.2~\mu$rad.

In the first step, the setup is tested using two successive 777 Bragg reflections
from Si crystals, Si(777)$\times$Si(777), as shown schematically in
Fig.~\ref{fig11}(a). The angular dependence of the Bragg reflection is
measured by rocking the angular orientation of the second crystal and
thus changing the angle of incidence of x-rays to the diffracting
atomic planes. The angular reflection profile shown in
Fig.~\ref{fig11}(a$^{\prime}$) features an angular width of
$\Delta\theta_{\mathrm{a}} = 2.7~\mu$rad, which is close to the
expected $\simeq 2~\mu$rad. The additional
broadening is attributed to a large horizontal spread of x-rays at a
bending magnet beamline. The appropriate spectral bandwidth is
calculated to be $\Delta E_{\mathrm{a}} = E \Delta\theta_{\mathrm{a}}
/\tan(\theta_{\ind{777}})\simeq 11$~meV. The angular width
$\Delta\theta_{\mathrm{a}}$ and the spectral width $\Delta
E_{\mathrm{a}}$ represent the angular and spectral resolution of the
setup with the reference Si(777) crystal.

In the next step, we used a similar nearly non-dispersive
double-crystal Bragg reflection arrangement, Si(777)$\times$C(800), as
shown schematically in Fig.~\ref{fig11}(b), to measure the angular
width of the single 800 Bragg reflection from one crystal plate of the
channel-cut crystal. The reflection profile shown in
Fig.~\ref{fig11}(b$^{\prime}$) on a logarithmic scale features an
angular width of $\Delta\theta_{\mathrm{b}} = 7.8~\mu$rad, which is
about $\simeq 2.5~\mu$rad broader than the expected 
value. The broadening is attributed to imperfections in the diamond
crystal. The appropriate Bragg reflection spectral width calculated
from the differential Bragg's law is $\Delta E_{\mathrm{b}} \simeq
31$~meV, which should be compared to a 20-meV single-reflection
spectral width.

In the final step, we used two successive Bragg reflections from both
plates of the (800) channel-cut crystal, as shown schematically in
Fig.~\ref{fig11}(c), to measure the combined angular Bragg reflection
width of the channel-cut crystal. Compared to the single-reflection
profile in Fig.~\ref{fig11}(b$^{\prime}$), the double-reflection
profile of the channel-cut crystal, shown in
Fig.~\ref{fig11}(c$^{\prime}$) on a logarithmic scale, features
steeper tails in the angular profile and a narrower angular width of
$\Delta\theta_{\mathrm{bc}} = 7.1~\mu$rad.  The appropriate Bragg
reflection spectral width calculated from the differential Bragg's law
is $\Delta E_{\mathrm{a}} = \simeq 28$~meV, which is broader than the
expected 17-meV (see Fig.~\ref{fig110}). The
double-reflection peak reflectivity drops by a factor of $\simeq 1.8$
compared to the single-reflection reflectivity in
Fig.~\ref{fig11}(b$^{\prime}$).  We attribute the decrease of the
double-reflection reflectivity to the angular misorientation of the
crystal plates.

Despite the deviations from the theoretical predictions, the
manufactured diamond channel-cut crystal is performing within 
expectations. The experimental data demonstrate the feasibility of
using channel-cut diamond crystals to monochromatize x-ray beams at cutting-edge XFELs.

\section{Conclusions and Outlook}
\label{conclusions}

We have demonstrated the feasibility of diamond channel-cut crystals
designed to function as high-heat-load, beam-multiplexing, narrow-band,
mechanically-stable x-ray monochromators for high-power x-ray beams at cutting-edge,
high-repetition-rate XFEL facilities.  Laser machining techniques were
used to manufacture complex 3D-structures in the channel-cut
crystals. The crystals were designed for use as x-ray
monochromators for 14.4~keV x-rays, corresponding to the $^{57}$Fe
nuclear resonance, and for 12.4~keV x-rays, corresponding to the
$^{45}$Sc nuclear resonance. The channel-cut crystals were
characterized by x-ray rocking curve imaging (topography) and by
measurement the angular and energy widths of double Bragg reflections.
The studied channel-cut crystals perform close to theoretical
expectations.  However, reflection curves are somewhat broadened, and
the channel-cut reflecting plates are misoriented by a few
microradians because of crystal defects, which further reduces the double-reflection crystal reflectivity.

High crystal quality in the initial diamond stone used in
fabricating the channel-cut crystal is essential for the proper
functioning of this diamond monochromator.

Along with applications as monochromators, other applications of
the diamond channel-cut crystals are anticipated.  In particular,
because diamond crystals feature very high x-ray reflectivity, they can be used as
efficient four-bounce or six-bounce channel-cut polarizers and analyzers to
achieve a very high degree of x-ray polarization purity \cite{BMS16}.

\section{Acknowledgments} 

Work at Argonne National Laboratory was supported by the
U.S. Department of Energy, Office of Science, Office of Basic Energy
Sciences, under contract DE-AC02- 06CH11357.  Diamond crystal growth,
quality checks and laser machining at FSBI TISNCM were performed on
a Shared-Equipment User Facility ``Research of Nanostructured, Carbon
and Superhard Materials''.

%\bibliography{/home/shvydko/laTeXdocs/bibliography/mybib}

\end{document}